\documentclass [preprint,12pt]{elsarticle}
\usepackage{amssymb,amsmath,graphicx,subfigure,breqn}
\usepackage{bm}
\usepackage{geometry}
\journal{Journal of Mathematical Chemistry}

\begin{document}
\title{When an oscillating  center in an open system undergoes power law decay}

\author{Sandip Saha }
\address{S N Bose National Centre For Basic Sciences\\Block-JD, 
Sector-III, Salt Lake, Kolkata-700106, India.}

\author{Gautam Gangopadhyay}

\address{S N Bose National Centre For Basic Sciences\\Block-JD, 
Sector-III, Salt Lake, Kolkata-700106, India.}

\ead{gautam@bose.res.in}

\date{\today}

\begin{abstract}
We have probed the condition of periodic oscillation in a class of two variable  nonlinear dynamical open systems modeled with Lienard-Levinson-Smith(LLS) equation which can be  a limit cycle, center or a very  slowly decaying  center type oscillation. Using a variety of examples of open systems like Glycolytic oscillator, Lotka-Volterra(L-V) model, a  generalised  van der Pol oscillator and   a time delayed nonlinear feedback oscillation  as a non-autonomous system, each of which contains a family of periodic orbits, we have solved LLS systems in terms of  a  multi-scale perturbation theory  using Krylov-Bogoliubov(K-B) method and it is utilised to characterise the size and shape of the  limit cycle and center as well as the  approach to their steady state  dynamics. We have shown the condition when  the average scaled radius of a center undergoes  a  power law decay  with exponent $\frac{1}{2}$. 
\end{abstract}
%\keywords{ 
%Isochronous oscillator,  Multiscale perturbation theory, Delay model, Limit cycle,  Power-law decay, K-B method, Lienard-Levinson-Smith(LLS) system}
\maketitle
\newpage
\section{Introduction}

Dynamical systems\cite{nayfeh,birkhoff,slross,ruelle,strogatz} capable of having isochronous oscillations\cite{len2,len3,sarkar,cologero} are very important from the point of view of modelling real-world systems\cite{len2,len3,len4} which exhibit self-sustained oscillation\cite{bender,murraynld,murraymbio,sabatini}. 
%Isochronicity is a widely studied subject not only for its relation with stability  and bifurcation theory\cite{bender,murraynld,murraymbio,sabatini} but also in the study of limit cycles\cite{len4} and oscillation with a center\cite{len2,len3}. 
The chemical oscillations\cite{prigogine,epstein,grayscott,kuramoto} are also of immense importance in biological world to maintain a cyclic steady state e.g., glycolytic oscillations\cite{glyscott,goldbook,gly2,gly3}, calcium  oscillations\cite{cal}, cell division\cite{cel}, circadian oscillation\cite{cir} and others\cite{epstein}. Although a lot of work has been performed in finding ways to determine if a system has a limit cycle, surprisingly a little is known about how to find this and still it remains a highly active area of research\cite{epstein,goldbook, strogatz,sarkar}. To obtain   the nonlinear dynamical features of a periodic orbit the general trend is  to resort to a geometrical approach coupled with tools of analysis\cite{nayfeh,birkhoff,slross,ruelle}. Recently RG analysis\cite{chen1,chen2} is heavily used to probe the multi-scale oscillation in the nonlinear system. Here a class of arbitrary autonomous kinetic equations in two variables  are cast into the form of a LLS oscillator\cite{len2,len3,len4,len0,len3.5} characterised by  the nonlinear  forcing and damping coefficients which can provide a unified approach to many problems concerning the existence %and the uniqueness 
of limit cycle and center.

In this work our focus is to characterize the properties of periodic orbits to distinguish among center and slowly decaying center type oscillation using an approximate solution of  a class of two variable equations from a  multi-scale perturbation theory by Krylov-Bogoliubov(K-B) approach\cite{slross}. Although there is  a condition which can distinguish limit cycle and center by using a constant part of damping coefficient in LLS equation but still  explicit  dynamical behaviour of their differences are not very well understood. We have  shown the   condition in which a center undergoes a slowly decaying orbit and their long time behaviour. Here in addition to the usual geometric disposition of periodic orbits with the  shape and size in the steady state  we have explored their asymptotic dynamics in terms of the energy consumption per cycle to distinguish  these  types of periodic orbits which  can appear in diverse situations.

In section 2, we have formulated the problem in terms of LLS oscillator. In section 3 taking examples of various systems we have studied the dynamical consequences of limit cycle, center and slowly decaying center type  in (a) Glycolytic oscillator, (b) L-V model, (c) van der Pol type oscillator and  (d) time delayed non-linear feedback oscillator.  In section 4, we have  studied all the above examples analytically and numerically and the main dynamical features of limit cycle, center and slowly decaying center type  are summarized in a table. In later section we have explored the source of slow decay of the center. The paper is concluded in section 6.

\section{Approximate solution  of LLS system: Description of the problem}
\label{sec 2}

Let us consider a two dimensional set of  equations for open system, % In this section we have solved Linenard equation using a multiscale perturbation theory where there exists some periodicity of the system and nonlinearity is weak. We consider a system of differential equations 
$$\frac{dx}{dt} = a_0+a_1x+a_2y+f(x,y),$$
\begin{equation}
\frac{dy}{dt} = b_0+b_1x+b_2y+g(x,y),
\label{eq1}
\end{equation}
where $x$ and $y$ are populations of two intermediate species of a dynamical process with $a_i$ and $b_i$ are all real parameters expressed in terms of the kinetic constants for all $i=0,1,2$ with $f(x,y)$ and $g(x,y)$ are  nonlinear functions for which  $x_s$ and $y_s$ are the steady state values.

Then  we define a new pair of variables, $(z,u)$ as
  $$u=\alpha_0+\alpha_1x+\alpha_2y ,$$
\begin{equation}
z=\beta_0+\beta_1x+\beta_2y, 
\label{eq2}
\end{equation}
where $\alpha_0,\alpha_1,\alpha_2$ and $\beta_0,\beta_1,\beta_2$
are constants  expressed in terms of $a_i$ and $b_i$.
%From the inverse transform of equation (\ref{eq2}), we can easily obtain the expressions of $x$ and $y$ as given by $$x=\frac{\beta_2 (u-\alpha_0)-\alpha_2 (z-\beta_0)}{\alpha_1 \beta_2-\beta_1 \alpha_2} =L(u,z),$$ and \begin{equation} \hspace{2cm}y=\frac{\beta_1 (u-\alpha_0)-\alpha_1 (z-\beta_0)}{\alpha_2 \beta_1-\beta_2 \alpha_1} =M(u,z). \end{equation}
Now considering $u$ and $z$ in such a way that, 
\begin{equation}
\frac{dz}{dt}=u,
\label{eq4}
\end{equation}
one can obtain an equation of $\ddot{z}$ from the equation of $u$. Now, using the steady state value $z_s=x_s+y_s$ one can find the LLS oscillator\cite{len4,len2,len3,strogatz,len0} for deviation from the stationary point from $z$ i.e. $\xi(=z-z_s)$ as

\begin{align}
\ddot{\xi}+F(\xi,\dot{\xi}) \dot{\xi}+G(\xi)&=0
\label{}
\end{align}
where, $F(\xi,\dot{\xi})$ is the damping function and $G(\xi)=\omega^2 \xi+O(\xi)$, is  an odd polynomial. Let us take $m= \lvert F(0,0) \rvert$ and rescaling the damping force function by $F_1 (\xi,\dot{\xi})$ such that $F(\xi,\dot{\xi})=m F_1 (\xi,\dot{\xi})$,
 LLS equation can be rewritten as

\begin{align}
\ddot{\xi}+m F_1 (\xi,\dot{\xi}) \dot{\xi}+G(\xi)=0.
\label{eq3}
\end{align}

An open system where oscillation is not straight forwardly obvious one can cast the 2D equations into LLS oscillator form which is amenable to multiscale perturbation analysis.
Note that the nature of the defining quantity of limit cycle and center is given by $F_1 (0,0)=j$, say. For limit cycle solution, $j=-1$, for  asymptotic solution, $j=+1$, and  $j=0$ if the nature of the solution of the system is center which means for the center case there is no time independent damping part. In both center and slowly decaying center type orbits one can find, $j=0$, so center and slowly decaying center type orbits can not be distinguished from the LLS equation form. From the linear stability analysis also one can not distinguish center from slowly decaying center type  cases as, $ F(0,0)=-2$ (Real part of eigenvalue) $=0$. 
The question is what is the source of this slow decay. 

Now rescaling the time variable $t$ by $\tau$ with $\tau = \omega t$ and $\xi(t)$ changing to $Z (\tau)$, one can obtain the form of a weakly nonlinear oscillator 
\begin{align}
\ddot{Z}(\tau)+\epsilon h(Z(\tau),\dot{Z}(\tau))+Z(\tau)=0,
\label{bkform}
\end{align}
where $\epsilon=\frac{m}{\omega^2}$ and $h(Z,\dot{Z})$  may contain  nonlinear damping term in LLS oscillator or explicitly time dependent terms for nonautonomous system such as time delayed oscillator and the control parameter $\epsilon$ must  lie in between 0 and 1 and more so if $0<m\ll \omega^2$, to have valid perturbative expansion.

%Now the above equation  becomes 
%\begin{align}
%\dot{Z}&=Y \nonumber \\
%\dot{Y}&=-Z -\epsilon h(Z,\dot{Z}),
%\label{eqoriginal}
%\end{align}
%where for  $\epsilon = 0$, the above system reduces to a simple harmonic oscillator with natural frequency 1 and the solution  looks like in the form $Z(\tau)=r \hspace{0.1 cm} \cos (\tau+\phi)$ and $Y(\tau)=-r \hspace{0.1 cm} \sin (\tau+\phi)$,  with $r=\sqrt{Z^2 +Y^2}$ and $\phi=-\tau+ tan^{-1} (- \frac{Y}{Z})$, where $r$ and $\phi$ are amplitude and phase, respectively with purely circular orbit. But for $\epsilon \neq 0$, the damping term is present and for small $\epsilon$ all orbits nearly become circular with periodicity $2 \pi$ in the scaled variable, $Z(\tau)$.

%Now making the amplitude $r$ and the phase $\phi$ slowly varying  time dependent to get an idea about the solutions of the above system, we consider, $Z(\tau)=r(\tau) \hspace{0.1 cm} \cos (\tau+\phi(\tau))$ and $Y(\tau)=-r(\tau) \hspace{0.1 cm} \sin (\tau+\phi(\tau))$  with $r(\tau)=\sqrt{Z^2 +Y^2}$ and $\phi(\tau)=-\tau+ tan^{-1} (- \frac{Y}{Z})$. Then one can obtain $\dot{r}(\tau)=\epsilon h \sin (\tau+\phi(\tau))$ and $\dot{\phi}(\tau)=\frac{\epsilon h}{r(\tau)} \cos (\tau+\phi(\tau))$ i.e. the time derivative of time dependent amplitude and phase are of $O(\epsilon)$.

Applying K-B with a running average $\overline{U}(\tau)= \frac{\varsigma}{2 \pi} \int_{0}^{\frac{2 \pi}{\varsigma}} U(s) ds$ over each cycle with $\varsigma$ as the natural frequency of the system\eqref{bkform}, %It is observed that $\dot{\overline{U}}=\overline{\dot{U}}$ from the fundamental theorem of calculus.
one can obtain,
\begin{align}
\dot{\overline{r}} &= \langle \epsilon \hspace{.15cm} h(Z,\dot{Z})\hspace{.15cm} \sin (\tau+\phi(\tau)) \rangle_\tau = \varphi_1 (\overline{r},\overline{\phi})\nonumber\\
\dot{\overline{\phi}} &= \langle \frac{\epsilon \hspace{.15cm} h(Z,\dot{Z})\hspace{.15cm}}{r(\tau)} \cos (\tau+\phi(\tau))\rangle_\tau = \varphi_2 (\overline{r},\overline{\phi}).
\label{eq7}
\end{align}
where $Z(\tau)=r(\tau) \hspace{0.1 cm} \cos (\tau+\phi(\tau))$.   Since $\dot{r}(\tau)$ and $\dot{\phi}(\tau)$ are of $O(\epsilon)$ then we may set the perturbation on $r$ and $\phi$ over one cycle as, $r(\tau) =\overline{r}+O(\epsilon)$ and  $\phi(\tau) =\overline{\phi}+O(\epsilon)$. %, where $\overline{r}$ and $\overline{\phi}$ are not exactly constant, they are very weakly $\tau$-dependent so that the error can be negligible.
%Finally one can obtain approximate equation from the above,
%\begin{align}
%\dot{\overline{r}} &= \varphi_1 (\overline{r},\overline{\phi}) \nonumber\\
%\dot{\overline{\phi}} &= \varphi_2 (\overline{r},\overline{\phi}),
%\label{eq10}
%\end{align}
%neglecting $O(\epsilon^2)$ terms where 
The functions $\varphi_1$ and $\varphi_2$ can be obtained from the explicit form of $h$ for the particular cases in the next section. The above system is in  a coupled form but most of the cases one can get decoupled set of equations so called amplitude and phase equations otherwise it will be quite harder to solve which is as good as the original system of equations; so it needs  to be solved by further approximation to get the amplitude equation.

From K-B approach one can define the system energy, $E= \frac{Z^2+\dot{Z}^2}{2}$ along with the energy consumption per cycle, $\Delta E =\frac{d}{d\tau}(\pi \overline{r}^2)$, where $O(\epsilon^2)$ terms are neglected.

\section{Some open systems and their comparative generic features}

Here we have studied a few prototypical examples to study the dynamical nature of limit cycle, center and slowly decaying center type periodic orbits. The first class of examples are from autonomous kinetic processes which are important in biology and chemistry, namely glycolytic oscillator\cite{glyscott,gly2,gly3,gly1}, L-V model\cite{epstein,goldbook, strogatz,sarkar} and a slightly generalized version of van der Pol type oscillator\cite{limiso,goldbook, strogatz} which can be converted to Leinard form. Next we have considered a nonautonomous system, a delay induced feedback model of nonlinear oscillator with a cubic nonlinearity. In the delay model we can find both limit cycle, center and slowly decaying center type oscillators in different parameter range.%  and its approximate solution is not very straightforward otherwise specially to obtain the bifurcation point.
  
\subsection{Glycolytic Oscillator}

The glycolytic oscillator\cite{glyscott,gly2,gly3,gly1} is found in the glycolysis by yeast, which can be described by the  overall dynamics and biochemical properties  of its enzyme phospho-fructokinase. Kinetic properties are simplified by the  mathematical analysis of Selkov\cite{gly1} and related models\cite{glyscott}.
Basic equations of a glycolytic oscillation is given by

\begin{align}
  \dot{x}(t) &= -x+(a+x^2) y \nonumber\\
  \dot{y}(t) &= b-(a+x^2) y,
  \label{eq1}
\end{align}
where $x$ and $y$  are the intermediate species concentrations with ($x_s = b, y_s =\frac{b}{a+b^2}$) as the fixed point. 
Considering, the above system in LLS form\cite{limiso}  and using the above  procedure in section \ref{sec 2}, one can find the approximate  solution of the above equation as
\begin{align}
x(t) &= b+\omega \overline{r} \sin(\omega t+\overline{\phi}) \nonumber\\
y(t) &= \frac{b}{a+b^2}+\overline{r} \lbrace \cos(\omega t+\overline{\phi})-\omega \sin(\omega t+\overline{\phi})\rbrace
\end{align}
%using above expressions of $\overline{r}$ and $\overline{\phi}$.
with 
\begin{align}
\dot{\overline{r}} &= - \frac{\epsilon \overline{r}}{ 8} \left( \frac{3 \hspace{0.15cm} \omega^2}{m \hspace{0.15cm} } \hspace{0.15cm} \overline{r}^2+4 j\right) \nonumber\\
\dot{\overline{\phi}} &= \frac{\overline{r}^2}{8 \hspace{0.15cm}},
\label{eq10}
\end{align}
along with the approximate system energy $E=\frac{1}{2} \left[ \lbrace x(t)+y(t)-z_s\rbrace^2+ \frac{\lbrace b-x(t)\rbrace^2}{\omega^2} \right]$.
%
%where $O(\epsilon^2)$ terms are neglected.
%
%
%Therefore, we have the solution of the form
%\begin{align}
%Z &= \overline{r} \cos (\tau+\overline{\phi})+O(\epsilon) \nonumber\\
%Y &= - \overline{r} \sin (\tau+\overline{\phi})+O(\epsilon)
%\label{eq11}
%\end{align}
%where $\overline{r}$ and $\overline{\phi}$ can be obtained by solving the set of equations (\ref{eq10}) using the initial conditions $r_0$  and $\phi_0$. 
%
%Now some points about the initial conditions are to be noted here. If we choose $x(t=0)=x_0$ and $y(t=0)=y_0$ as an initial condition for the original system then $z(t=0)=z_0=x_0+y_0$ and $\dot{z}(t=0)=b-x_0$ $\Rightarrow$ $\xi(t=0)=x_0+y_0-z_s=A$ and $\dot{\xi}(t=0)=b-x_0=B$ and $\dot{\xi}(0)=0=B$ only when $x_0=b=x_s$. Also, initially at $t=0$ we have $\tau=0$, therefore one can find $Z(0)=\xi(0)=A$ and $Y(0)=\frac{1}{\omega} \dot{\xi}(0)=\frac{B}{\omega}$. Then, $r_0=$ initial value of $r=\frac{1}{\omega} \sqrt{\omega^2 A^2+B^2}$ and $\phi_0=$ initial value of $\phi=tan^{-1} \left(- \frac{B}{\omega A}\right)$.
%
Now three cases may arise as in this example, $j$ can assume any of its three values. The condition of having a stable limit cycle is given by $j=-1$, with the radius of the cycle $\frac{2 }{\omega }\sqrt{\frac{m}{3}}$ at $t\rightarrow \infty$. Next, the condition of having a stable fixed point is with $j=1$  and in the asymptotic limit radius goes to zero exponentially. And finally, the condition of having a decaying center type solution for $j=0$, where the radius decays  as a power law   with asymptotic expression, $\overline{r} \propto t^{-\frac{1}{2}}$.

\subsection{Lotka-Volterra Model}

Lotka-Volterra(L-V) equation is the basic prey predator model\cite{epstein,goldbook, strogatz,sarkar} which is known to have an oscillatory solution and this oscillation is shown to be have a center. The prey population is $x$ and the predator population is $y$ with the dynamical equation is given by,

\begin{align}
\dot{x}(t) &= \alpha x -\beta x y \nonumber\\
\dot{y}(t) &= -\gamma y+\delta x y,
\label{lveq}
\end{align}
with $\alpha,\beta,\gamma,\delta>0$. The two fixed points are ($x_s = 0, y_s =0$) and ($x_s = \frac{\gamma}{\delta}, y_s =\frac{\alpha}{\beta}$) where the first one is a saddle point and the later one gives a center  solution obtained from linear stability analysis. The LLS form of the above system \eqref{lveq} is given in Appendix. The approximate analytical solution takes the form,
\begin{align}
x(t) &= \frac{\gamma}{\delta}+\frac{\overline{r}}{ z_s \delta}\lbrace\gamma \cos(\omega t+\overline{\phi})-\omega \sin(\omega t+\overline{\phi}) \rbrace \nonumber\\
y(t) &= \frac{\alpha}{\beta}+\frac{\overline{r}}{ z_s \beta}\lbrace \alpha \cos(\omega t+\overline{\phi})+\omega \sin(\omega t+\overline{\phi}) \rbrace,
\end{align}
and $E=\frac{1}{2} \left[ \lbrace \delta x(t)+\beta y(t)-z_s\rbrace^2+ \frac{\lbrace \alpha \delta x(t)-\beta \gamma y(t)\rbrace^2}{\omega^2} \right]$. Here $\dot{\overline{r}} = 0$ and $\dot{\overline{\phi}} = 0$, indicates no correction in amplitude as well as phase are needed.

\subsection{A Generalised van der Pol Oscillator}

We consider here a little generalised form of van der Pol Oscillator\cite{limiso,goldbook,strogatz} instead of considering  $a=1$ which is traditionally used. This generalised system can provide both a limit cycle, and a slowly decaying center type solution by changing the  parameter,  $a$, unlike the traditional van der Pol oscillator with only a Limit cycle solution with nonzero value of $\epsilon_1$. 
Basic equations are given by,

\begin{align}
  \dot{x}(t) &= y \nonumber\\
  \dot{y}(t) &= - \epsilon_1 y (x^2-a^2)-\omega^2 x,
  \label{eq}
\end{align}
where $0<\epsilon_1 \ll 1$ is a  small perturbative constant and ($x_s = 0$, $y_s =0$) is the only fixed point.

Considering LLS form\cite{strogatz,limiso},
finally one can obtain the approximate analytical solution  of the form, 
\begin{align}
x(t) &=\overline{r} \cos(\omega t+\overline{\phi}) \nonumber\\
y(t) &= -\omega \overline{r} \sin(\omega t+\overline{\phi}),
\end{align}
 where,
\begin{align}
\dot{\overline{r}} &= - \frac{\epsilon \overline{r}}{8 a^2} \left( \overline{r}^2-4 a^2 \right) \nonumber\\
\dot{\overline{\phi}} &= 0,
\label{eq10}
\end{align}
and $E=\frac{1}{2} \left(x^2+\frac{y^2}{\omega^2}\right)$ with $\epsilon=\frac{\epsilon_1 a^2}{\omega}$.

For this system two cases can arise, a limit cycle and a slowly decaying center type solution. Condition of having a stable limit cycle for a real positive value of $a$, $\dot{\overline{r}}$ becomes, $\dot{\overline{r}} = - \frac{\epsilon \overline{r}}{8 a^2 } \left( \overline{r}^2-4 a^2  \right)$ which gives the radius of the limit cycle is $2a$ as $t\rightarrow \infty$. Again, the condition of having a slowly decaying center type solution can be obtained for $a=0$ which gives, $\dot{\overline{r}} = -\frac{\epsilon_1}{8\hspace{0.15cm}\omega  } \hspace{0.15cm} \overline{r}^3$ and it gives a power law decay as $t \rightarrow\infty$ with $\overline{r} \propto t^{-\frac{1}{2}}$.

\subsection{Time Delayed  Nonlinear Feedback Oscillator}

Many nonlinear dynamical systems in various scientific disciplines are influenced by the finite propagation time of signals in feedback loops modelled with a time delay\cite{len3.5,goto}. In some systems, such as lasers and electro mechanical models, a large variety of delays appear. We have provided a time delayed model to obtain both limit cycle, center and slowly decaying center type\cite{len3.5,limiso} oscillation for different range of parameters of the system. There is no general scheme to handle delay system using perturbation theory\cite{mickens} to study bifurcation or periodic orbit and characterisation  of the properties of oscillation in limit cycle, center and slowly decaying center type cases\cite{len3.5}.

First we consider  here a model of delay system where the oscillation is fed energy  through a delay term  and its total energy increases with time. We have introduced a linear damping term to make a center solution and then in presence of damping and delay when we introduce another quartic nonlinear term\cite{kuramoto,goto} in the potential it gives a limit cycle and slowly decaying center type solution by tuning the parameters.
The basic equations of the model are
\begin{align}
  \dot{x}(t) &= y(t) \nonumber\\
  \dot{y}(t) &= - \epsilon \lbrace x(t-t_d)+\dot{x}(a x^2+b)\rbrace-\omega^2 x(t),
  \label{eq}
\end{align}
where we have considered, $0<\epsilon \ll 1$ is a small perturbative parameter and ($0,0$) is the only fixed point. So, it becomes a delayed van der Pol system which induces energy into the system.  Using K-B averaging scheme one can have the analytical solution 
\begin{align}
x(t) &= \overline{r} \cos(\omega t+\overline{\phi}) \nonumber\\
y(t) &= - \omega \overline{r} \sin(\omega t+\overline{\phi}),
\end{align}
where,
\begin{align}
\dot{\overline{r}} &= -\frac{\epsilon \overline{r}}{8} \lbrace a \overline{r}^2-4\left(\frac{\sin \omega t_d}{\omega}-b\right)\rbrace, \nonumber\\
\dot{\overline{\phi}} &= \frac{\epsilon}{2 \omega} \cos(\omega t_d),
\label{eq10}
\end{align}
and $E =\frac{\omega^2 x^2+\dot{x}^2}{2}$ assuming  $0<t_d \ll 1$.

\section{Numerical Results for various open systems}

Here we have numerically explored the characteristics of periodic orbits of limit cycle, center and slowly decaying center type cases in various physical systems namely,  Glycolytic, Lokta-Volterra and van der Pol type oscillator  and a delayed nonlinear feedback oscillator. We have shown the validity of approximate amplitude equations in terms of the phase space and the role of phase space dynamics. In these model systems we have shown the characteristic features of  limit cycle, center and slowly decaying center type orbits through  their  asymptotic approach to steady state in terms of a scaled radius and average energy consumption per cycle.  

\begin{figure}
\begin{center}
\includegraphics[width=\textwidth]{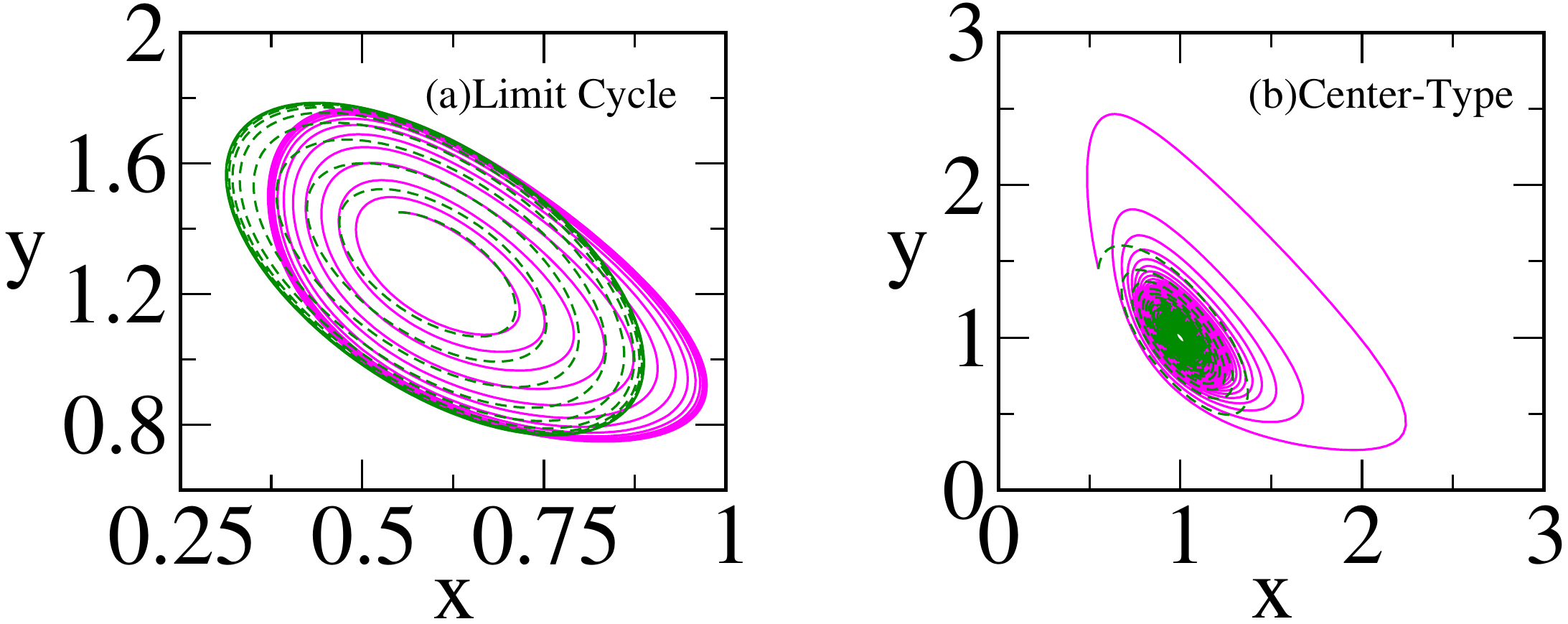}
\caption{\textbf{Glycolytic Oscillator:} (a) Limit cycle phase portrait with $a=0.11$ and $b=0.6$ and (b) slowly decaying center type phase portrait with $a=0$ and $b=1$, where dotted line indicates the numerical simulation of the approximate analytical solution and the solid one is the exact numerical solution of the system.}
\label{fig 1}
\end{center}
\end{figure}

As a starting example we consider Glycolytic Oscillator where for the different values of the constants, $a$, $b$  can provide limit cycle or a slowly decaying center type orbit. At first we  discuss for the limit cycle case in this system which arises for $a=0.11$, $b=0.6$ and then slowly decaying center type case with $a=0$, $b=1$ with the initial condition(IC) of $x=0.55$ and $y=1.45$.  Figure \ref{fig 1} shows the phase space portrait of (a) a stable limit cycle  and (b) slowly decaying center type oscillation where the dotted lines indicates the numerical simulation for the approximate solution and the solid one indicates the exact numerical solution. From the figure 1(a) and (b), it is clear that the approximate solution shows the same nature as the exact one and both gives the limit cycle solution except  a phase lag. Reason can be found in the  non-zero value of $\dot{\overline{\phi}}$ which brings a phase lag. 

\begin{figure}[h!]
\begin{center}
\includegraphics[width=\textwidth]{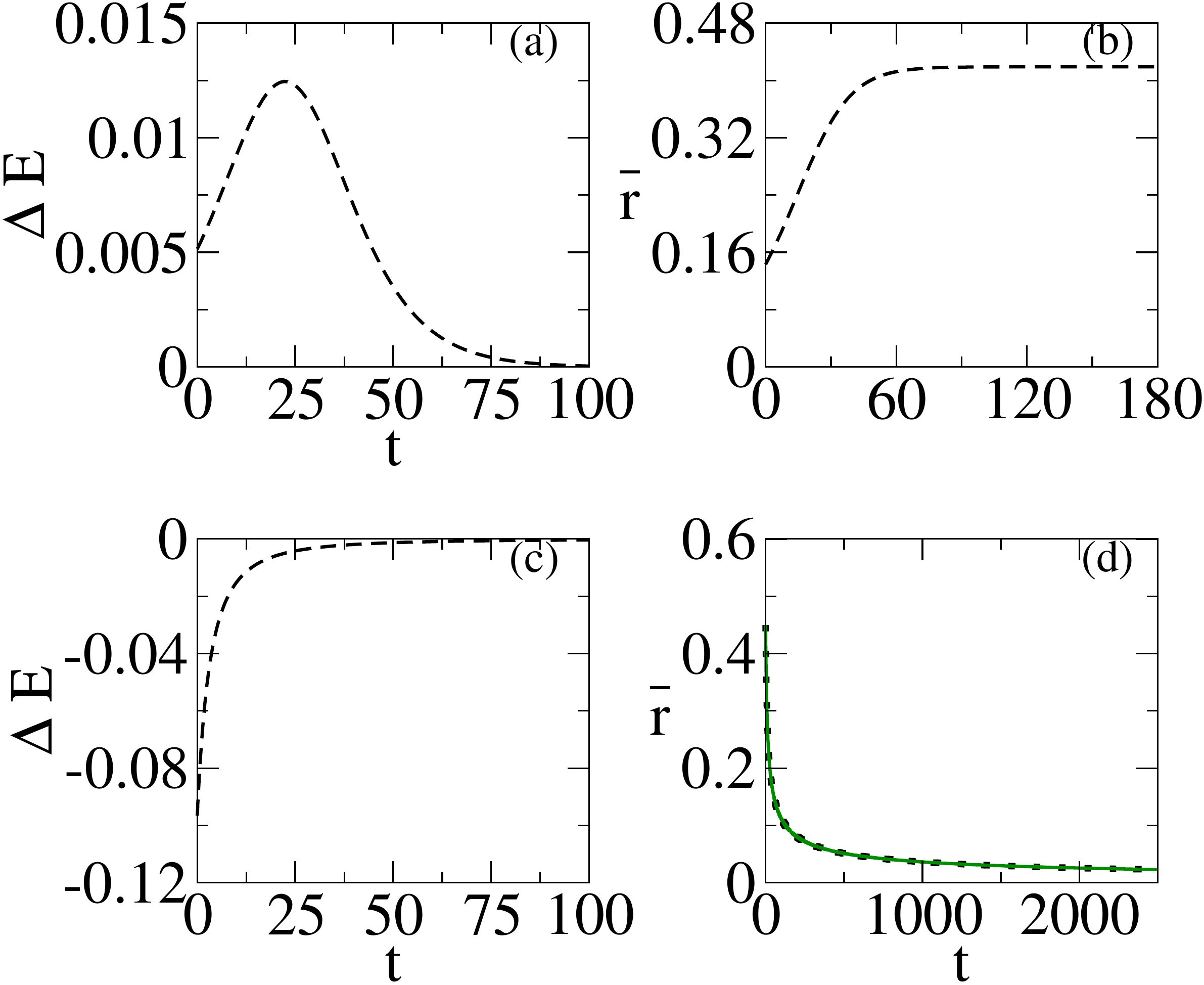}
\caption{\textbf{Glycolytic Oscillator:} For the limit cycle case(a-b) when the IC is inside the orbit: (a) energy consumption per cycle goes to zero in the steady state after passing through a maximum;  (b) scaled radius of the limit cycle is shown. For the case of slowly decaying center type:  (c)energy consumption per cycle starts with a very small negative value to reach zero as time increases (d) the scaled radius decreases with a power law decay where  the dotted one is the fitting curve.}
\label{fig 2}
\end{center}
\end{figure}

In \ref{fig 2}(a-b) we have shown the dynamics of the scaled radius and energy consumption per cycle of the limit cycle.  In \ref{fig 2}(c-d) similar features of the dynamics of  scaled radius and energy change per cycle of the slowly decaying center   are shown. From Figure \ref{fig 2}(a) and Figure \ref{fig 2}(c) we can say that in both  limit cycle and slowly decaying center type cases change of energy per cycle must be zero in the long time limit, however, for the limit cycle it is passed through a maximum as the IC is taken inside the limit-cycle. From Figure \ref{fig 2}(b) and Figure \ref{fig 2}(d) the scaled radius in limit cycle and slowly decaying center type cases approach to steady state in a different manner. For limit cycle it becomes a constant and for slowly decaying center it goes as power law decay as $\overline{r}(t)\approx t^{-\frac{1}{2}}$. Figure \ref{fig 2}(d) gives a power law fitted curve $r(t)=A_0 (1+A_1 t)^{-0.5}$ with $A_0=0.449517$ and $A_1=0.151655$.

\begin{figure}
\begin{center}
%\minipage{0.47 \textwidth}
%\begin{center}
\includegraphics[width=0.5\textwidth]{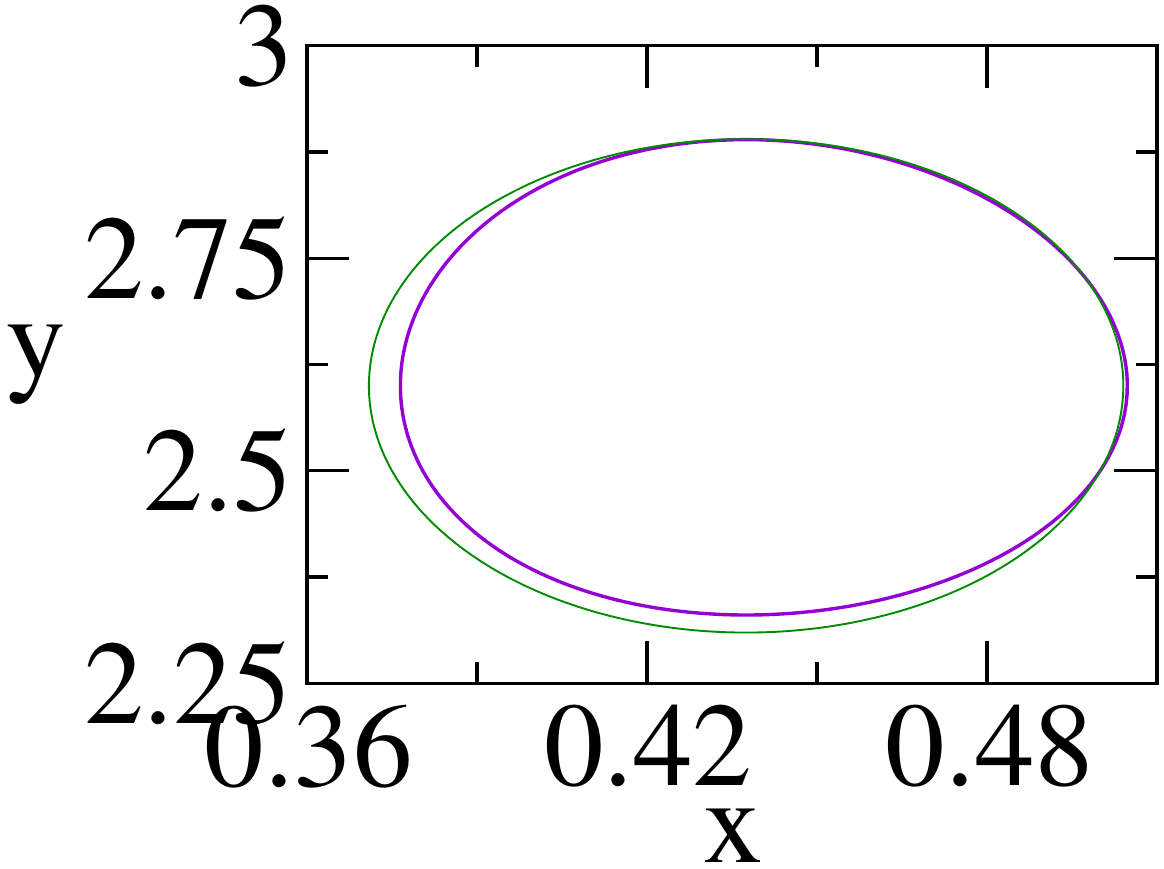}
\caption{\textbf{L-V:} Phase space graph of the  center with $\alpha=1.3, \beta=0.5, \gamma=0.7, \delta=1.6$ and $\epsilon=0.1$, where exact and approximate curves have no phase lag.}
\label{fig 3}
\end{center}
\end{figure}

In the next section we have analysed the L-V equation with  $\alpha=1.3, \beta=0.5, \gamma=0.7, \delta=1.6$ and $\epsilon=0.1$ with initial values	$x_0=0.5$ and $y_0=2.5$. 
Figure \ref{fig 3} shows a phase portrait for the case of  center of L-V system  with the exact numerical(solid) and approximate solution with amplitude equation(dashed).  The dynamics of the scaled radius is a constant from the initial time.  Average energy consumption per cycle is zero from the initial time for the case of center which is quite different from the case of limit cycle and slowly decaying center type orbit. The features of scaled radius and average energy per cycle for the Center is attributed to the fact that the  initial point is always on the orbit for the case of a center. 
Since here $\dot{\overline{\phi}}=0$, there does not exist any phase lag between the exact and approximate centers shown in Figure \ref{fig 3} which also reflects the energy change per cycle with time which is zero  as $\dot{\overline{r}}=0 \implies \overline{r}=r_0$ and depends on the initial value which defines the radius of the center.

\begin{figure}
\begin{center}
\includegraphics[width=\textwidth]{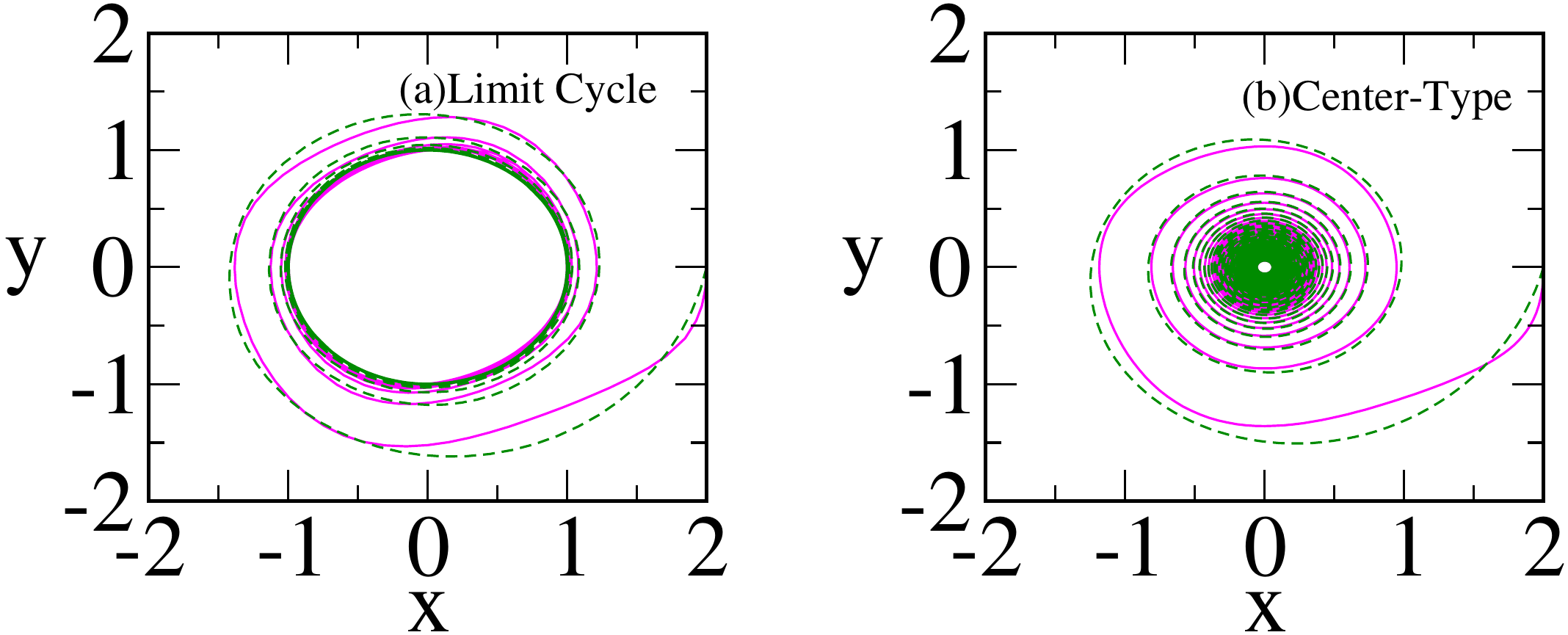}
\caption{\textbf{van der Pol type oscillator:} (a) Limit cycle phase portrait for $a=0.5$ and (b) slowly decaying center type phase portrait for $a=0$ where $\epsilon$ is fixed with $0.5$ in both cases. The dotted lines indicates the numerical simulation of the approximate analytical solution and the exact numerical solution of the system in the solid line.}
\label{fig 4}
\end{center}
\end{figure}

\begin{figure}
\begin{center}
\includegraphics[width=\textwidth]{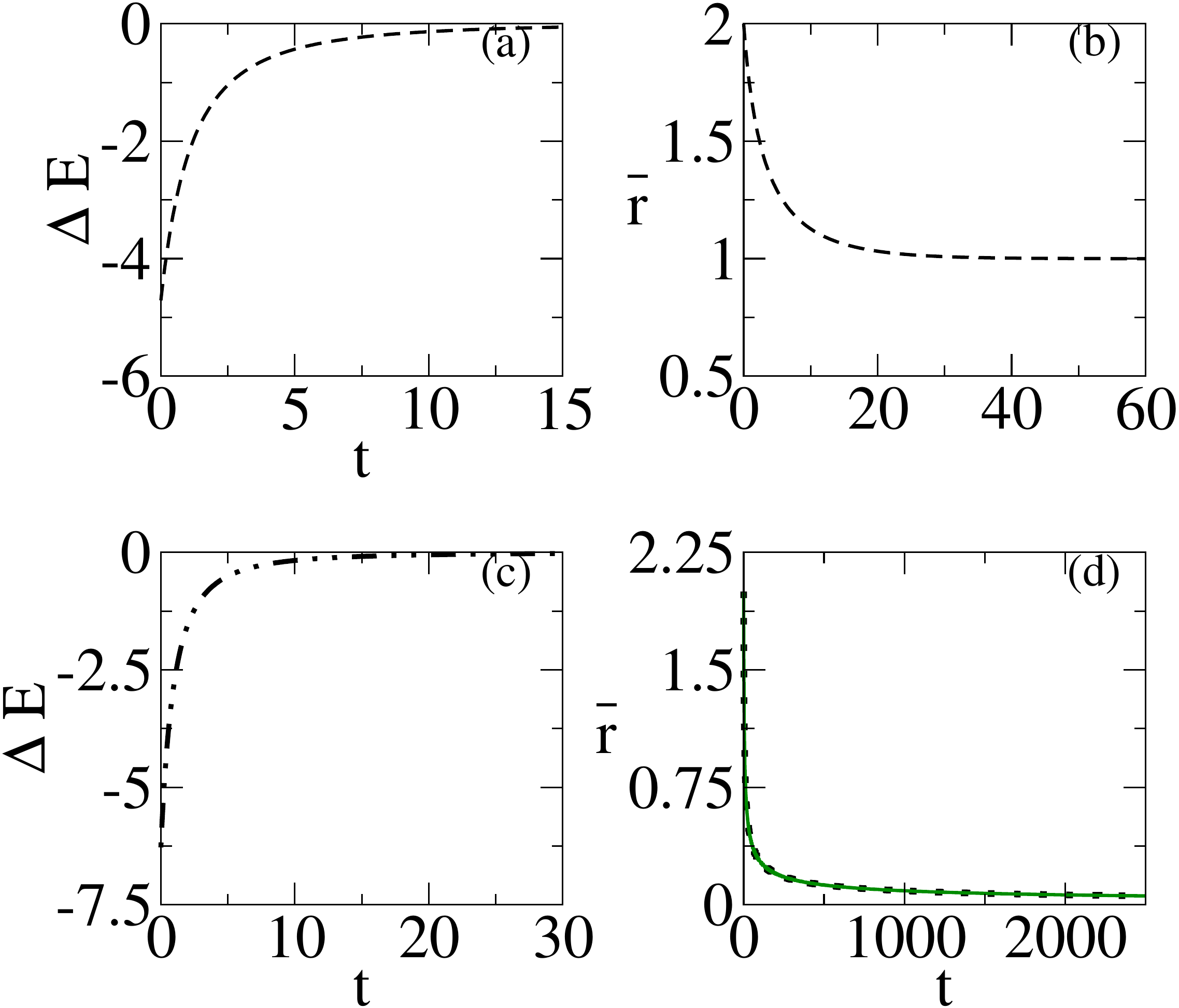}
\caption{\textbf{van der Pol type oscillator:} For the limit cycle case(a-b) when the IC is outside the orbit: (a) energy consumption per cycle starting from a negative value it goes to zero in the steady state;  (b) scaled radius of the limit cycle assumes a constant value in the steady state. For the case of slowly decaying center type:  (c)energy consumption per cycle starts with a very small negative value to reach zero as time increases (d)the scaled radius decreases with a power law decay where  the dotted one is the fitting curve.}
\label{fig 5}
\end{center}
\end{figure}

In the next section we have analysed a little generalized version of van der Pol equation where $a$ is a parameter unlike usual van der Pol case($a=1$)  with $a=0.5$ for the limit cycle and $a=0$ for slowly decaying center type case 
with $\epsilon=0.5$ and IC, ($x_0=2,y_0=0$)  is taken from outside of the cycle. %As we mentioned previously about the differences of initial points inside and out side of the cycle, the graphs shows the differences accordingly.
 Since here we find $\dot{\overline{\phi}}=0$ then there does not exist any phase lag between the two cycles where it is shown in Figure \ref{fig 4} phase portrait and in the Figure \ref{fig 5} shows scaled radius and the energy consumption per cycle with time.
In fig 5(a-b) shows the energy consumption per cycle and scaled radius with time as expected from the limit cycle case. Energy consumption per cycle would pass through a maximum if the IC would be inside the orbit as in the glycolytic case which is not shown in figure. In fig 5(c-d) it is the slowly decaying center type case which shows the energy consumption per cycle gives the same behaviour as in 5(a) but the scaled radius with time gives a power law decay with a fitted curve, $r(t)=A_0+A_1 (1+A_2 t)^{-0.5}$ with $A_0=0.000167043$, $A_1=1.99394$ and $A_2=0.0498935$.

\begin{figure}
\centering
\begin{subfigure}[]
\centering
\includegraphics[width=5.5cm]{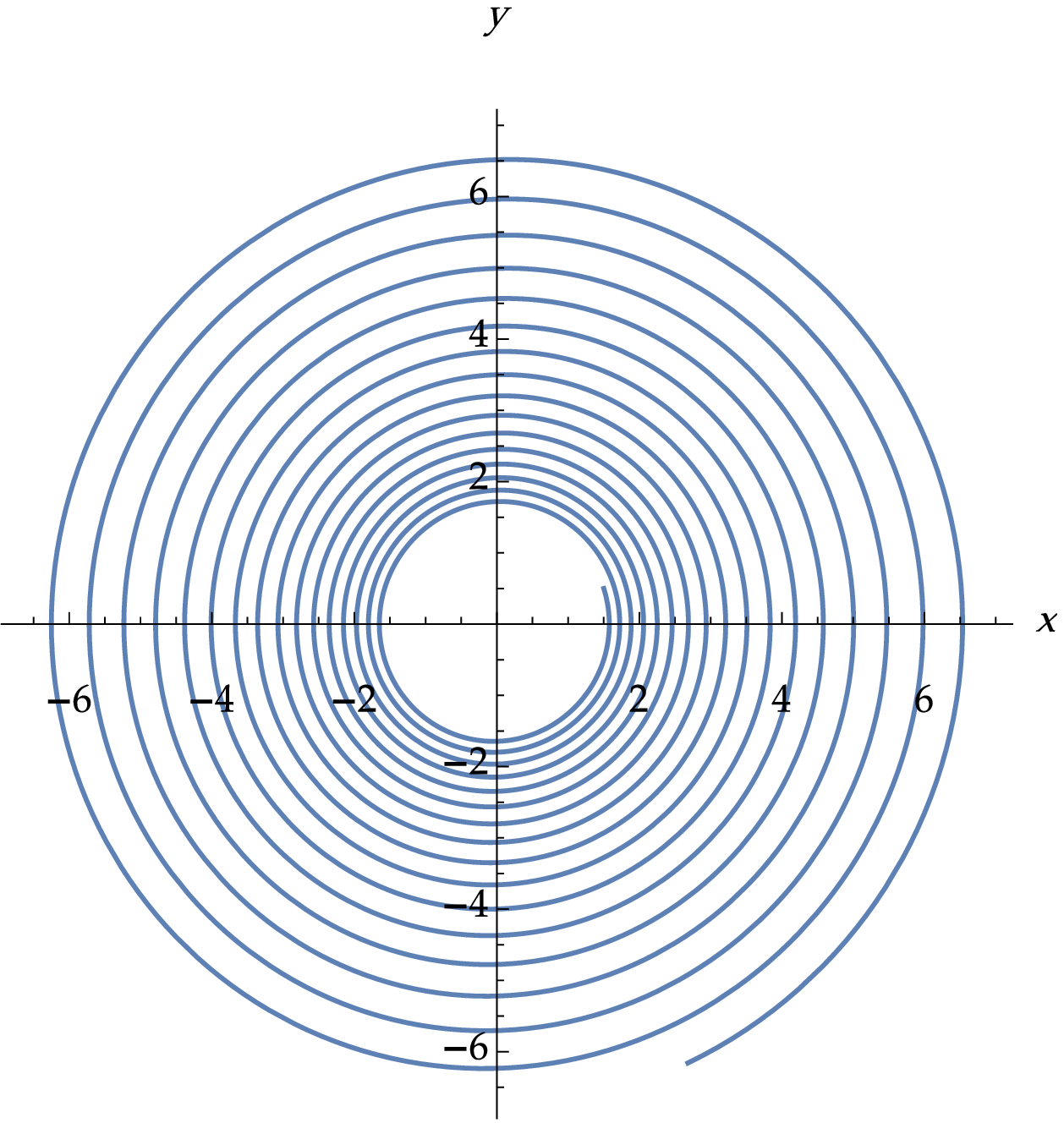}
%\caption{}
\end{subfigure}
\begin{subfigure}[]
\centering
\includegraphics[width=1.8 cm, angle =-30]{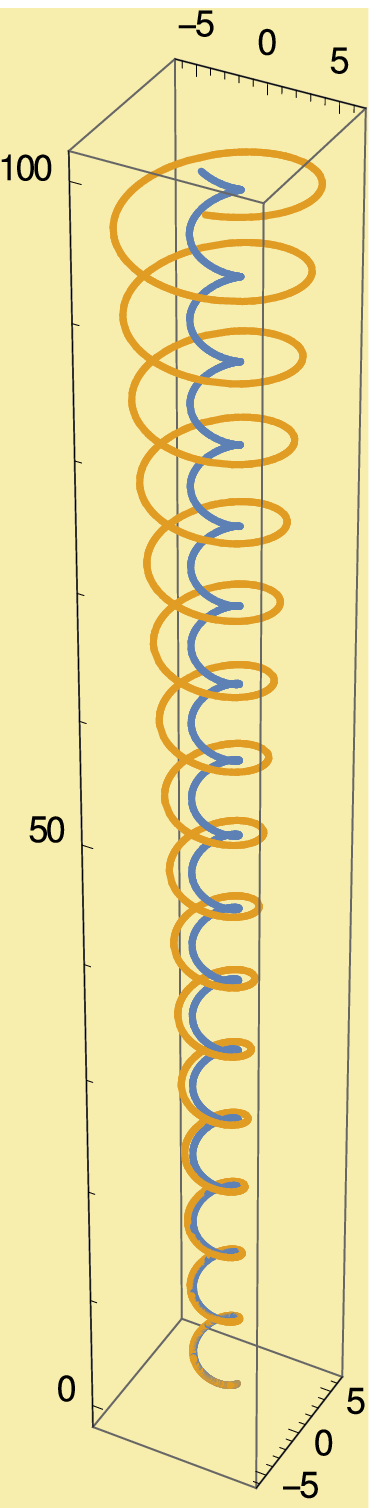}
%\caption{}
\end{subfigure}
\caption{\textbf{Time-Delayed System:} (a) Phase portrait of delay induced feedback oscillator with no damping and nonlinearity, $a=0,b=0$ with $t_d=0.623$ (directly calculated in Mathematica) (b) 3D phase space plot of delay (outer curve) in respect of non-delay (inner curve) with time.}
\label{fig 6}
\end{figure}

\begin{figure}
\begin{center}
%\minipage{0.45\textwidth}
\includegraphics[width=\textwidth]{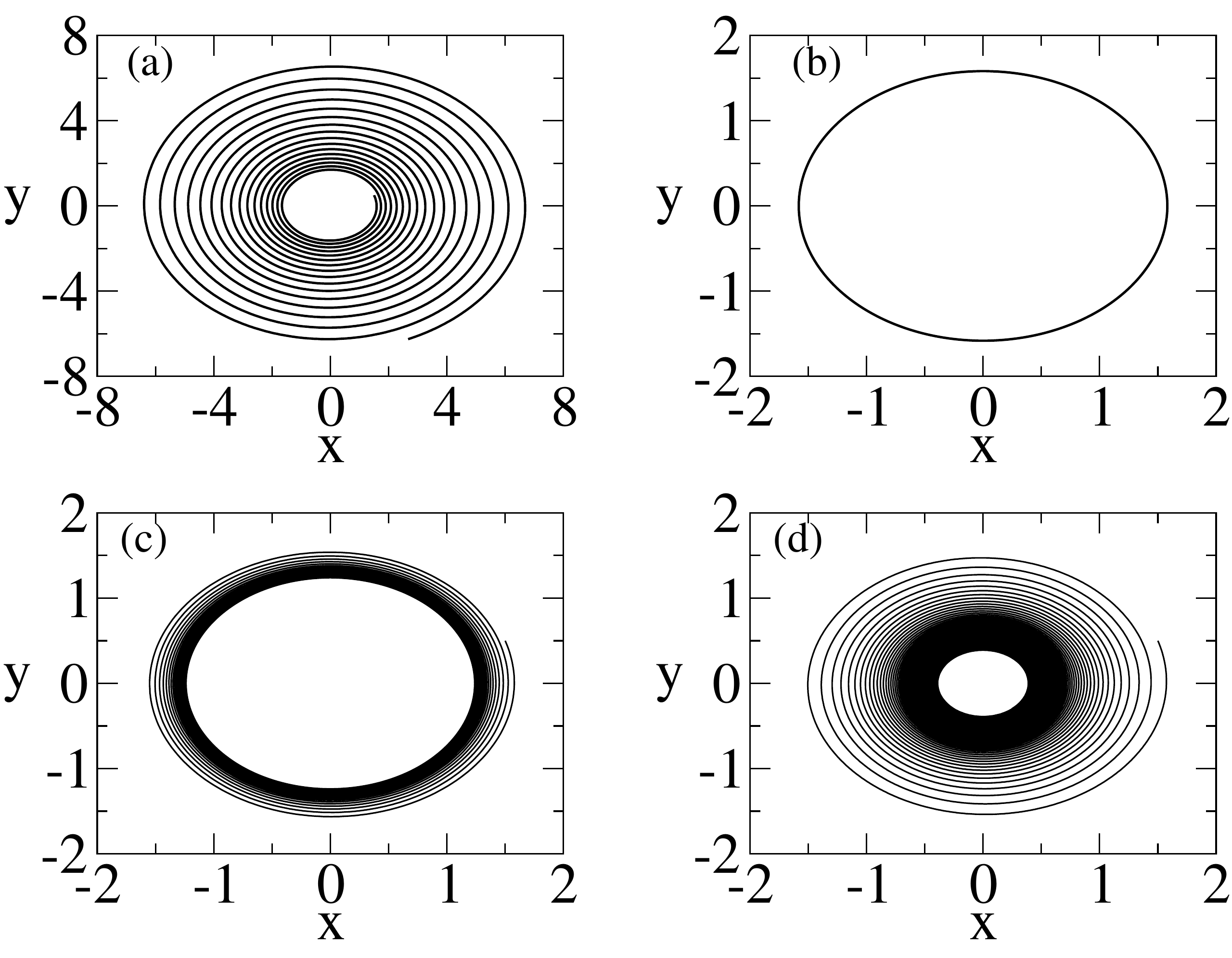}
\caption{\textbf{Time-Delayed System:} Phase space plots of the numerical simulation of the approximate amplitude equation with the same time delay for (a) $a=0,b=0$ one gets a feedback system with increasing phase space area, (b) a center with $a=0,b=\frac{Sin(\omega t_d)}{\omega}$, (c) a limit cycle with $a=1,b<\frac{Sin(\omega t_d)}{\omega}$ and (d) a slowly decaying center type orbit with $a=1,b=\frac{Sin(\omega t_d)}{\omega}$}
\label{fig 7}

\end{center}
\end{figure}

For the delay model with no nonlinearity and damping, i.e, $a=0,b=0$ is a simple feedback oscillator with continuously increasing energy in the system. The phase portraits are shown in fig \ref{fig 6} and fig \ref{fig 7} with a finite delay, $t_d=0.623$  
with taking the parameters,   $\epsilon=0.05$ and $\omega=1$ and the initial value $x_0=1.5$ and $y_0=0.5$. Figure \ref{fig 6}(b) shows the phase space effect of delay (outer curve) in respect of non-delay (inner curve) in 3D with time. 
 As in this case both $\dot{\overline{r}} \neq 0$ and $\dot{\overline{\phi}} \neq 0$ so there may exist a phase lag between the two phase space plots and the lag  increases with time delay.  
 
  In figure \ref{fig 7} we have shown the phase space curves in (a) feed back system with no damping and nonlinearity where area is increasing, \ref{fig 7}(b) a phase space for the center,  \ref{fig 7}(c) a limit cycle  and in \ref{fig 7}(d) a slowly decaying center type orbit for different parameters of $a$ and $b$.
  
\begin{figure}
\begin{center}
%\endminipage\hfill
%\minipage{0.45\textwidth}
\includegraphics[width=\textwidth]{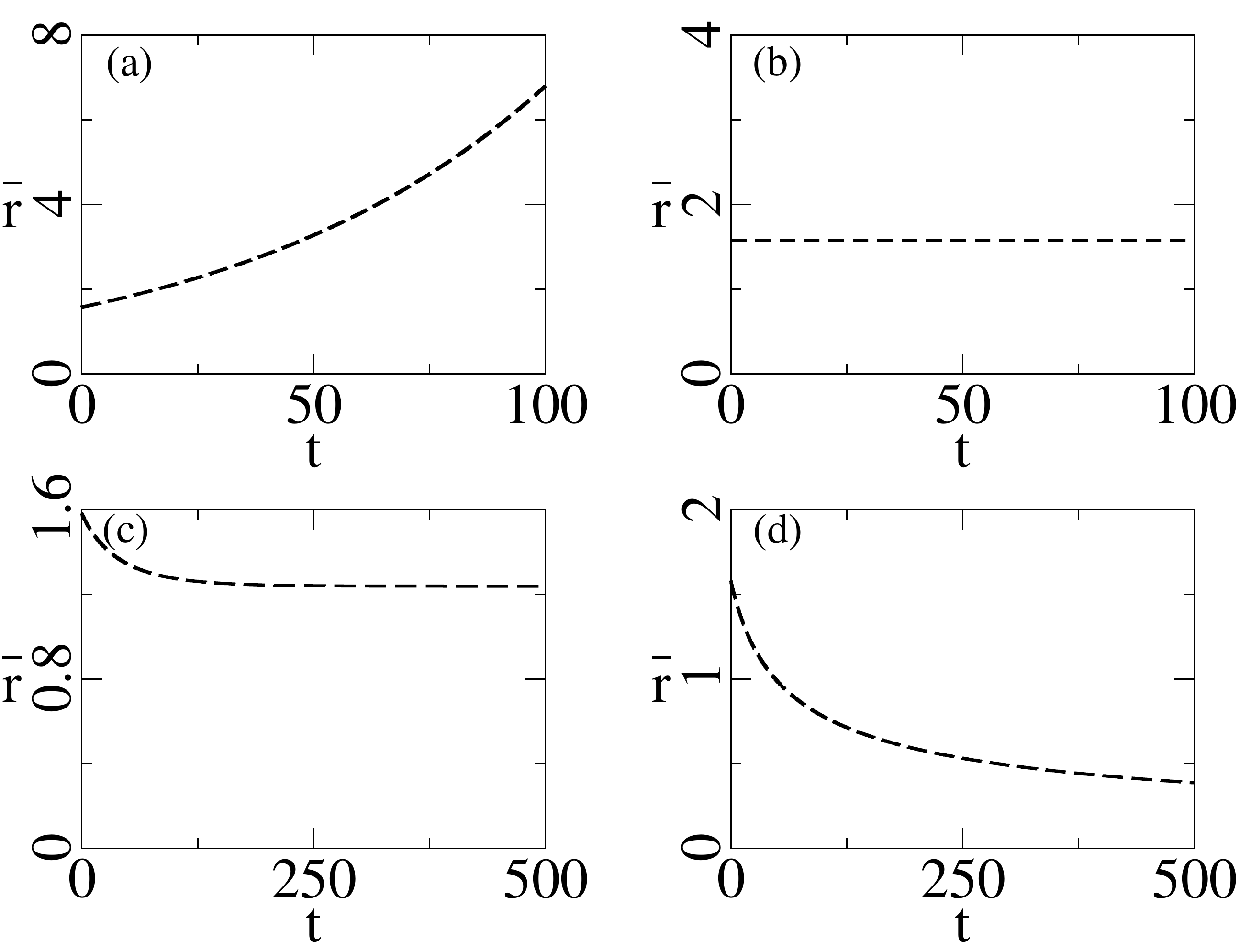}%td_r_vs_t.eps}
\caption{\textbf{Time-Delayed System:} Scaled  radius is shown as a function of time in (a) increasing exponentially for purely  feedback case, in (b)   a constant from the initial time for the Center,  in (c)  changes from its IC outside the cycle  to reach a constant corresponding to a limit cycle  and in (d)  decreases slowly with power law decay for the case of  a slowly decaying center type orbit. } 

\label{fig 8}
\end{center}
\end{figure}

 In figure \ref{fig 8} we have shown the time dependent nature of the scaled radius for different parameters of $a$ and $b$. In purely feedback system with no damping and nonlinearity almost exponentially increasing radius in (a), \ref{fig 8}(b) a radius is a constant from the initial time for the center,  \ref{fig 8}(c) radius changes from its inital value to reach a constant corresponding to a limit cycle  and in \ref{fig 8}(d) radius decreases slowly with power law decay for the case of  a slowly decaying center type orbit. 

\begin{figure}
%\minipage{0.45 \textwidth}
\begin{center}
\includegraphics[width=\textwidth]{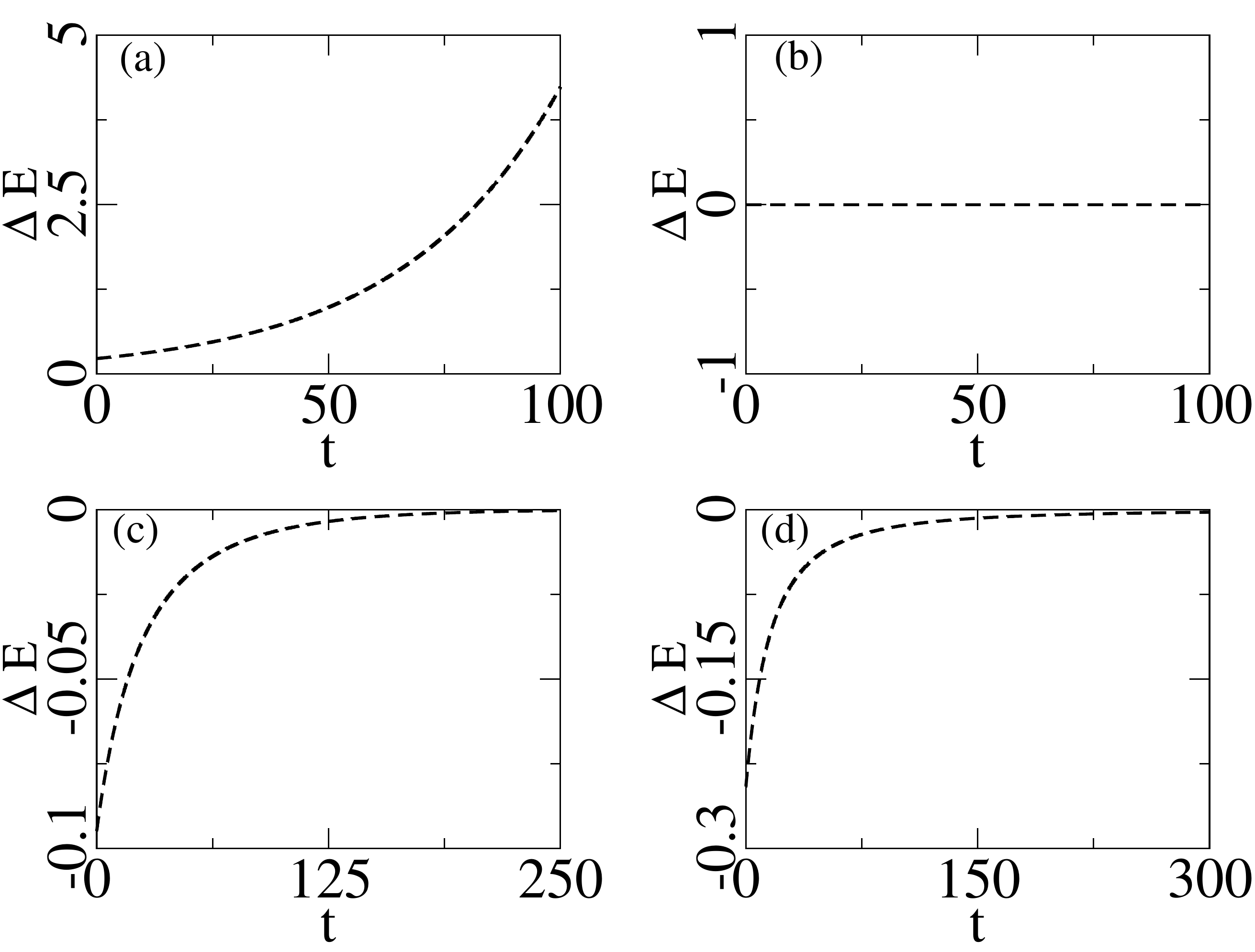}%td_Del_E_and_E_vs_t_E_is_red_curve_and_D_E_is_black.eps}
\caption{\textbf{Time-Delayed System:} The average energy consumption per cycle($\Delta E$) is shown as a function of time in (a) increasing exponentially for purely  feedback case, in (b)   a constant from the initial time for the center,  in (c)  changes from its IC out side the cycle to reach  zero corresponding to a limit cycle  and in (d)  goes to zero for the case of  a slowly decaying center type orbit.}
\label{fig 9}
\end{center}
\end{figure}

In figure \ref{fig 9} we have shown the time dependent nature of the average energy consumption per cycle($\Delta E$)  for different parameters of $a$ and $b$. For feedback system with no damping and nonlinearity almost exponentially increasing $\Delta E$ in (a), \ref{fig 9}(b) $\Delta E$ is a constant from the initial time for the Center,  \ref{fig 9}(c) $\Delta E$ changes from its inital value to its vanishing value corresponding to a limit cycle  and in \ref{fig 9}(d) $\Delta E$ goes to zero from its initial value for a slowly decaying center type orbit.

The results of the numerical exploration are summarized in  Table-1 to illustrate the dynamical features of limit cycle, center and slowly decaying center type orbits. The multiscale perturbation theory is adopted here from the Lecture notes of Strogatz which is based on the K-B averaging method. It is applied for the LLS system and a non-autonomous system of delayed nonlinear feedback  model. The scaled  amplitude equation generates all the results of shape, size of the limit cycle, center and slowly decaying center type orbits  almost exactly except in few limit cycle cases a phase lag is found which needs the equation of phase also to be solved simultaneously. The amplitude equation can also be utilised as a stability criteria where the nature of the periodic orbit can be explicitly assigned. This method is giving comparable result with RG method which we have not shown here but checked analytically and numerically. However, it can not give correct result for strongly nonlinear cases where the restoring force in the LLS equation is nonlinear.
\newpage
\begin{table}
\textbf{\Huge{ }}
\begin{center}
\includegraphics[width=16 cm]{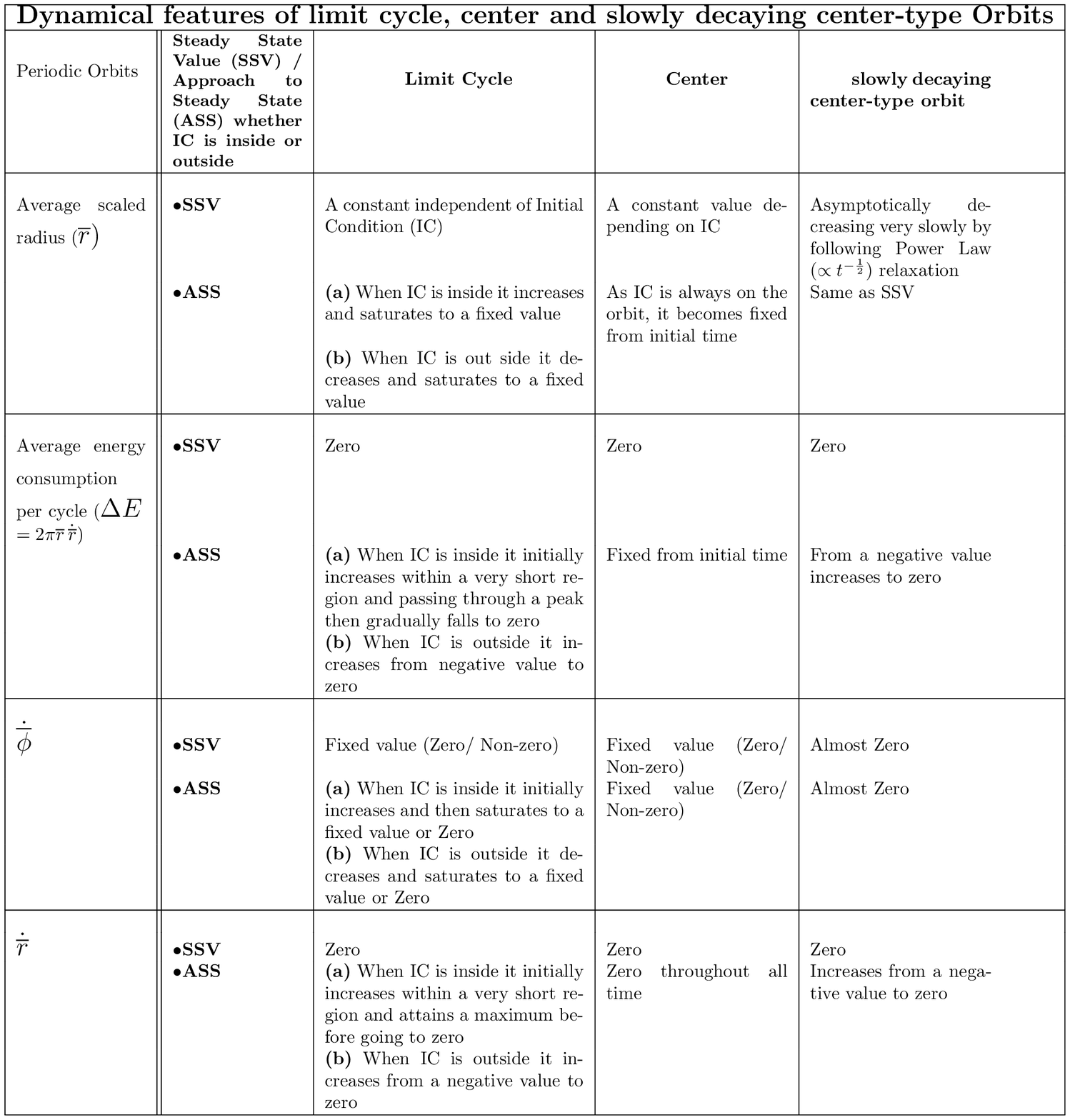}%\vspace{-4.3 cm}
\caption{Dynamical features of limit cycle, center and decaying center-type orbits}
\end{center}
\end{table}

\section{Source of Power Law Decay}

We have shown in  various two dimensional open systems where the slowly decaying center  undergoes a power law decay with exponent $\frac{1}{2}$. The question is how a center undergoes a power law decay? The source of power law decay can be traced  in the nonlinear damping  function in the LLS system.
 
It is very straight forward to  show that the restoring force will not affect the amplitude equation. To show that we  restrict ourselves to the case of LLS equation where the $F(\xi,\dot{\xi})$ and $G(\xi)$ are the polynomial functions of $\xi$ and $\dot{\xi}$. It is well known that linear functional forms of $F$ and $G$ preclude the existence of center. This can be readily seen by considering the typical examples, e.g., a harmonic oscillator or a weakly nonlinear oscillator with a potential $\frac{1}{2} \omega^2 x^2+\frac{1}{3} \lambda x^4$, $0 < \lambda <1$ or a Lotka-Volterra model, where one can encounter a center.  

In LLS system $F(\xi,\dot{\xi})$ and $G(\xi)$ can be polynomial function of $\xi$ and $\dot{\xi}$ and depending on the even-odd  properties of $F(\xi,\dot{\xi})$ and $G(\xi)$ amplitude and phase of the oscillation affect the center and limit cycle  of the system. In what follows we employ K-B method of averaging to show that the characteristic even and odd powers of polynomials play crucial role in determining the behaviour of the associated amplitude and phase equations.

Unlike a harmonic oscillator which has a center solution, for a nonlinear damping case,  $\ddot{x}+\omega^2 x=-\lambda x^3, 0 <\lambda \ll 1$,  one can find a decaying center solution\cite{len3.5}. Next, considering both  damping and non-linear restoring force, a typical example of Lotka-Volterra system with LLS form(see Appendix), $\ddot{x}+\epsilon_1 (b_1 x+b_2 \dot{x}) \dot{x}+\omega^2 x+\epsilon_1 b_3 x^2=0, 0 < \epsilon_1 \ll 1$, gives a center solution. Here, the damping force function is of linear order and also odd as $F(\xi,\dot{\xi}) \neq F(-\xi,-\dot{\xi})$. From the various examples, it is not clear when a center undergoes power law decay. So we would like to introduce a little more general scenario when it appears.

To begin with we consider polynomial functions of  $F(\xi,\dot{\xi})$ and $G(\xi)$ with upto cubic power   of $\dot{\xi}$ and $\xi$ in the following reduced forms,
\begin{align}
F(\xi,\dot{\xi}) &=-[A_{01}+A_{11} \xi+A_{21} \xi^2+A_{31} \xi^3 +A_{02} \dot{\xi}+A_{12} \xi \dot{\xi}+A_{22} \xi^2 \dot{\xi}+A_{32} \xi^3 \dot{\xi} \nonumber\\
&+A_{03} \dot{\xi}^{2}+A_{13} \xi \dot{\xi}^{2}+A_{23} \xi^2 \dot{\xi}^{2}+A_{33} \xi^3 \dot{\xi}^{2}], \nonumber\\
G(\xi) &=-[A_{10} \xi+A_{20} \xi^{2}+A_{30} \xi^{3}].
\end{align}
Let us take $|F(0,0)|=\sigma \in \mathbb{R}^+$, an arbitrary constant with $F(\xi,\dot{\xi})=\sigma F_{\sigma} (\xi,\dot{\xi})$. Then the LLS equation can be rewritten as
\begin{align}
\ddot{\xi}+\sigma F_{\sigma} (\xi,\dot{\xi}) \dot{\xi}+G(\xi)=0.
\label{eq3}
\end{align}
Therefore the final equation takes the form of a non-linear oscillator after rescaling  $t$ by $\tau$ taking, $\omega t \rightarrow \tau$ as
\begin{align}
\ddot{Z}(\tau)+\epsilon h(Z(\tau),\dot{Z}(\tau))+Z(\tau)=0,
\label{14}
\end{align}
where, $0<\epsilon=\frac{\sigma}{\omega^2}\ll 1$, 
$\omega^2=-A_{10}>0$ and $Z(\tau) \equiv \xi(t)$ and $\dot{Z}(\tau) \equiv \dot{\xi}(t)$. Equation \eqref{14} is amenable to  averaging with  K-B method which gives 
\begin{dmath}
h(Z,\dot{Z})=-\left[\lbrace B_{01}+ B_{11} Z+B_{21} Z^2+B_{31} Z^3 + B_{02} \omega \dot{Z} + B_{12} Z \omega \dot{Z}+ B_{22} Z^2 \omega \dot{Z} + B_{32} Z^3 \omega \dot{Z}+ B_{03} \omega^2 \dot{Z}^{2}+B_{13} Z \omega^2 \dot{Z}^{2}+ B_{23} Z^2 \omega^2 \dot{Z}^{2}+B_{33} Z^3 \omega^2 \dot{Z}^{2} \rbrace \omega \dot{Z}+ B_{20} Z^{2}+B_{30} Z^{3}\right], 
\end{dmath}
where $B_{ij}=\frac{A_{ij}}{\sigma}$, $i,j=1,2,3$ are the corresponding indices. The amplitude and phase equations are obtained as,
\begin{align}
\dot{\overline{r}} &= \frac{\epsilon  \omega \overline{r} }{16} \lbrace\overline{r}^2 \left(B_{23} \overline{r}^2 \omega ^2+6 B_{03} \omega ^2+2 B_{21}\right)+8 B_{01}\rbrace+O(\epsilon^2), \nonumber\\
\dot{\overline{\phi}} &= -\frac{\epsilon \overline{r}^2}{16} \left(B_{32} \overline{r}^2 \omega ^2+2 B_{12} \omega ^2+6 B_{30}\right)+O(\epsilon^2).
\label{eq10}
\end{align}
Now from a detailed analysis  of the amplitude equation  for $\dot{\overline{r}}$, it is apparent that only even elements of $F(\xi,\dot{\xi})$  appears but none of the elements of $G(\xi)$ is present due to the vanishing value of the averages of $\sin^{\mu} \cos^{\nu}$ terms with $\mu=1$ and $\nu \in \mathbb{Z}$. 
The non-zero averages arise  only when $\mu,\nu$ both are even i.e. $\mu=2 \eta_1, \nu=2  \eta_2; \eta_1,\eta_2 \in \mathbb{Z}$. 
The power law solution of amplitude with $t^{-\frac{1}{2}}$  can  appear from equation\eqref{eq10} only  when the right hand side contains $r^3$ term which means when $A_{23}$ or $B_{23}$ terms should be absent and there must be nonzero positive value of any one of the  terms,  $A_{03}$ and $A_{21}$ should be present.

\section{Conclusions}
By suitably adopting K-B averaging method  in multi-scale perturbation theory for a periodic system here we have provided the solution of a class of two variable open systems through LLS form.  The  approximate K-B solution is shown to be almost exact for calculating  physical properties with a set of diverse examples, namely,   Glycolytic oscillator, L-V system, a generalized van der Pol oscillator   and  a  time delayed nonlinear feedback oscillator.  To characterize a slowly decaying center type oscillator one can find that even when the constant part of damping force in LLS equation vanishes i.e, $ F(0,0)=0$, the center may undergo a slow decay and asymptotically gives a stable focus and the radius of the center undergoes a power-law decay. Here we have investigated about the source of this power law decay and in this context  we have compared the asymptotic dynamics of the  limit cycle, centre and slow decay of centre-type orbit in various open systems and a generic feature of all these systems are explored. 

The condition of isochronicity which is usually defined as the amplitude independent  period of the orbit  is utilized to characterize the dynamical features of limit cycle, center and slowly decaying center type orbits as pointed below.  While the shape of the orbits  are shown in phase space of actual variables, the size of the orbit is studied by introducing an average scaled radius variable and asymptotic approach of the orbits to steady state can be  understood from the scaled radius and energy consumption per cycle.
\begin{enumerate}
\item From the approximate solution  we can find the  limit cycle, center or slowly decaying center type motion almost exactly just by using the single variable equation of the radius, so called amplitude equation. However, to get rid of the slight phase lag in the limit cycle case one needs to solve the equation of phase variable coupled with radius variable simultaneously.
\item The energy consumption per cycle $\Delta E$, of the limit cycle vanishes at the steady state depending on the position of the initial condition which can be inside or outside of the cycle in  a particular way:  $(a)$ when the initial condition is inside the cycle, it increases initially to a maximum in a very short time and then goes to zero, $(b)$ when the initial condition is outside the cycle, 
$\Delta E$ increases from a negative value to zero. For the case of a center quite distinctly it is zero from the initial time upto the steady state. For a slowly decaying center, $\Delta E$ increases from a negative value to zero.
\item Most interesting feature about the difference in center and slowly decaying center type oscillation, which is indistinguishable from the LLS equation form, i.e., $F(0,0)=0$, the slowly decaying center type case reveals a power law decay of the radius, $t^{-\frac{1}{2}}$ asymptotically unlike the center where the radius becomes almost constant from initial time, however, $\Delta E$ for slowly decaying center type oscillation arrives at zero in a finite time.
\end{enumerate}

\section{Appendix}
\textbf{Casting Lotka-Volterra system into LLS form}: Let us set $z=\delta x+\beta y$ which gives $\dot{z}=\alpha \delta x-\beta \gamma y=u$ $\implies$ $x=\frac{\dot{z}+\gamma z}{(\alpha+\gamma)\delta}$ and $y=\frac{-\dot{z}+\alpha z}{(\alpha+\gamma)\beta}$.
After taking $t$ derivative upon $\dot{z}$ one can have,
\begin{align*}
\ddot{z} &=(\alpha-\gamma)\dot{z}+\alpha \gamma z +\frac{\dot{z}^2}{\alpha+\gamma}+\frac{\gamma - \alpha}{\alpha+\gamma} z \dot{z}-\frac{\alpha \gamma}{\alpha+\gamma} z^2.
\end{align*}
As the fixed point, $(0,0)$ gives a saddle  solution, choosing the remaining non-zero fixed point for further investigations and after taking perturbation $z=\xi+z_s$ around the fixed point $z_s=\alpha+\gamma=\delta x_s+\beta y_s \neq 0$, one can get the LLS form with
%
%\begin{align}
%\ddot{\xi}+F(\xi,\dot{\xi}) \dot{\xi}+G(\xi)=0
%\label{eq2}%%
%\end{align}
%where, 
$F(\xi,\dot{\xi})=a_1 \xi +a_2 \dot{\xi}$ % is linear and having weak nonlinearity in the LLS equation 
where $a_1=\frac{\alpha-\gamma}{\alpha+\gamma}$ and $a_2= - \frac{1}{\alpha+\gamma}$. It is to be noted that $G(\xi)$ contains nonlinearity with $G(\xi)=\omega^2 \xi+a_3 \xi^2$ where $\omega=\sqrt{\alpha \gamma}=Im(\lambda)$($+ve$ sense) and $a_3=\frac{\alpha \gamma}{\alpha+\gamma}$. After introducing a very small parameter $\epsilon_1$ (say) in the constants, $a_i, b_i$  such that $a_i=\epsilon_1 b_i, i=1,2,3$ the above equation reduces to $\ddot{\xi}+\epsilon_1 (b_1 \xi+b_2 \dot{\xi}) \dot{\xi}+\omega^2 \xi+\epsilon_1 b_3 \xi^2=0$, and after rescaling time it reduces to $\ddot{Z}(\tau)+\epsilon h(Z (\tau),\dot{Z}(\tau))+Z (\tau)=0$ with $h=(k_1 Z + k_2 \dot{Z}) \dot{Z}+k_3 Z^2$ where $k_1=\omega b_1$, $k_2=\omega^2 b_2$,$k_3=b_3$, and $\epsilon=\frac{\epsilon_1}{\omega^2}$. % and since we are analysing it in the weak nonlinear case then $\epsilon$ must be set in between 0 and 1 i.e. $0<\epsilon\ll1$ i.e. $0<\epsilon_1\ll\omega^2$. 
We have considered $0<\epsilon\ll1$, which means $0<\epsilon_1\ll\omega^2=\alpha \gamma \le 1$ $\implies \alpha \le \frac{1}{\gamma}$, since $\alpha,\gamma>0$.

%****************************************************************************
\noindent
{\bf Acknowledgement}

\noindent
{ Sandip Saha acknowledges RGNF, UGC, India for the partial financial support. We thank Prof. D S Ray for useful discussion.}

\end{document}